\documentclass[12pt]{article}

\usepackage{comment}
\usepackage{mathtools}
\usepackage{url}
\usepackage{amssymb,amsmath}
\usepackage{graphicx}
\usepackage{subfigure,epsfig}
\title{A Note on Circular Arc Online Coloring using First Fit}

\author{Paraskevas V. Lekeas\\\small{6807 N Keeler Ave, Lincolnwood}\\\small{IL 60712, USA}\\\small{plekeas@gmail.com}}
\date{}
\begin{document}

\maketitle

\begin{abstract} {In \cite{Raman} using a column construction technique it is proved that every interval graph can be colored online with First Fit with at most $8w(G)$ colors, where $w(G)$ is the size of the maximum clique of $G$.  Since the column construction can not be adapted to circular arc graphs we give a different proof to establish an upper bound of $9w(G)$ for online coloring a circular arc graph $G$ with the First Fit algorithm.}
\end{abstract}

\section{Introduction}

A Circular arc graph is the intersection graph of arcs of a circle \cite{Golumbic, Tucker}, while Interval graph is a graph that has an intersection model consisting of intervals on the real line \cite{Golumbic}. Circular arc graphs are somehow a generalization of Interval graphs \cite{Damaschke,ISGCI}. The problem of online coloring an interval graph using the FirstFit (\textit{FF}) algorithm is the following: The intervals of the graph appear one at a time and for each interval \textit{FF} must assign the least available color\footnote{Colors are supposed to be positive integers $1,2,3,\cdots$ .}. A lot of research has been conducted to bound the maximum number of colors that \textit{FF} will use \cite{Raman,Gyarfas,Kierstead}. In his Ph.D. R. Raman \cite{Raman} proved that for every interval graph $G$ with maximum clique size $w(G)$, \textit{FF} needs at most $8w(G)$ colors to color online $G$. Now moving our attention to online coloring circular arc graphs, to the best of our knowledge the performance of \textit{FF} has not been studied. In \cite{Marathe} a heuristic algorithm for online coloring a circular arc graph is given and in \cite{Slusarek} an algorithm that partially uses \textit{FF} for online coloring a circular arc graph is used. In this note we prove that if $G$ is a circular arc graph with maximum clique size $w(G)$ then \textit{FF} needs at most $9w(G)$ colors to color online $G$. Although this result is very close to Raman's theorem for interval graphs, the construction proposed in \cite{Raman} can not be adapted to circular arc graphs. The reason is that in circular arc graphs a unique natural ordering of the arcs - crucial for the column construction in Raman's theorem - does not exist. This fact motivated us to find a different way to bound the number of colors used by $FF$ to color online a circular arc graph. In the rest of the note in section 2 we describe the problem of online coloring\footnote{From now on online coloring refers to online coloring with \textit{FF}.} an interval graph proving a crucial Lemma. In section 3 we describe the problem of online coloring a circular arc graph for the case where the circular arc graph has a minimum clique size of 1. In the last section we discuss the general case where the circular arc graph $G$ contains a minimum clique of size $K \leq w(G)$.

\section{Interval graph online coloring using $FF$}

Assume $G'$ is an interval graph with $m+1$ intervals and $w(G')$ the maximum clique size of $G'$. Let $\Sigma=<\delta_1,\delta_2,...,\delta_{m+1}>$ be a sequence (ordered $(m+1)$-tuple) denoting the order of appearance of the intervals of $G'$ in a valid online coloring with $FF$. Assume that the $FF$ algorithm, for every sequence $\Sigma$, builds another sequence of positive integers, $X=<\chi_1,\chi_2,...,\chi_{m+1}>$, which denotes what color was assigned to every corresponding interval of $\Sigma$. Let us call $X$ the chromatic sequence of $\Sigma$. From Raman's theorem \cite{Raman} and by definition of the $FF$ algorithm, it holds that, $1 \leq \chi_i \leq 8w(G'), \forall \chi_i \in X, \, \, 1 \leq i \leq m+1$.

\begin{figure}[htb]
\centering 
\includegraphics[scale=0.5]{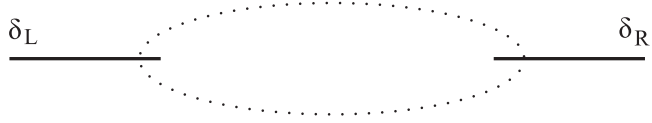}
\caption{The interval graph $G'$. Inside the eclipse there exist $m-1$ intervals. Intervals $\delta_L$ and $\delta_R$ do not overlap.}
\label{G'}
\end{figure}

Let $G'$ be an interval graph as before and suppose $\delta_L$ is the leftmost interval of $G'$. Let $\delta_R$ also be the rightmost interval of $G'$. Assume for now that $\delta_L$ and $\delta_R$ are unique and also assume that these two intervals do not intersect in $G'$ (we will later see what this assumption means for our problem). In Figure \ref{G'} we can see $G'$.

In an online coloring of $G'$, from all the possible, $(m+1)!$ in number, sequences $\Sigma$, we choose all these that have the following characteristic: The intervals $\delta_L$ and $\delta_R$ appear consecutively, i.e. one after the other. Let $\Sigma_1=<\delta_1,\delta_2,\cdots,\delta_i,\delta_L,\delta_R,\delta_{i+3},\cdots,\delta_{m+1}>$ be such a sequence and let $X_1=<\chi_1,\chi_2,\cdots,\chi_i,\chi_{\delta_L},\chi_{\delta_R},\chi_{i+3},\cdots,\chi_{m+1}>$ be the chromatic sequence that $FF$ produces from $\Sigma_1$. As we mentioned earlier, no element of $X_1$ is bigger than $8w(G')$.

Assume now that during the online process of coloring $\Sigma_1$, $FF$ assigns $\chi_{\delta_L}=L$ and $\chi_{\delta_R}=R$, where $L,R$ are colors $\leq 8\omega(G')$ (it is obvious that there exists such a possible evolution of the online process). We wish to find out how the chromatic sequence $X_1$ would be affected if someone pauses the online coloring process immediately after the $i$-th step is completed, assigns $\chi_{\delta_L}=\Delta$ and $\chi_{\delta_R}=\Delta$, where $\Delta=8w(G')+1$, and lets it roll again from step $i+3$. Of course the part of $X_1$ that would get affected would be the subsequence $<\chi_{i+3},\chi_{i+4},...,\chi_{m+1}>$ and this is because for all the previous intervals a valid color has been assigned and these assignments are permanent according to the definition of $FF$. Let $<\chi'_{i+3},\chi'_{i+4},...,\chi'_{m+1}>$ be the chromatic subsequence that $FF$ would produce from $(i+3)$-th step (including) and onwards if we didn't pause the process after the end of the $i$-th step (let us call this process the non-pause process). We have now the following

~\

\textbf{Lemma 1}: $\chi_j \leq \chi'_j \leq 8w(G')$, $\forall j, \, \, i+3 \leq j \leq m+1$.

\textbf{Proof}: Assume that $X'_1=<\chi'_{1},\cdots,\chi'_i,\chi'_{\delta_L},\chi'_{\delta_R},\chi'_{i+3},\cdots,\chi'_j,\cdots,\chi'_{m+1}>$ is the chromatic sequence in the non-pause process and 

$X_1=<\chi_{1},\cdots,\chi_i,\chi_{\delta_L},\chi_{\delta_R},\chi_{i+3},\cdots,\chi_j,\cdots,\chi_{m+1}>$ is the chromatic sequence in the pause process. The proof is by induction on $j$.

\underline{Base}: $j=i+3$. Assume that in the non-pause process we had that $\chi'_{\delta_L}=L$, $\chi'_{\delta_R}=R$ and $\chi'_{i+3}=\Pi$. We have the following cases:

\begin{itemize}

\item $\delta_{i+3}$ does not overlap with either $\delta_L$ or $\delta_R$. Since $\delta_{i+3}$ received in the non-pause process color $\Pi$, this means that $\Pi$ was the first (smaller) available color. In the pause process now the overlapping constraints for $\delta_{i+3}$ are exactly the same with the non-pause process and since $\chi_1=\chi'_1, \cdots, \chi_i=\chi'_i$, $\Pi$ is still the smaller color available for $\delta_{i+3}$, so $\chi_{i+3} = \chi'_{i+3}$ and thus $\chi_{i+3} \leq \chi'_{i+3}$ as desired. 

\item $\delta_{i+3}$ overlaps either with $\delta_L$, or with $\delta_R$, or with both, then in any case we have the following. 

\begin{itemize}

\item if in the non-pause process it is the case that $\Pi < R$ and $\Pi < L$, then after the pause we will have that $\chi_{i+3}=\chi'_{i+3}=\Pi$. This is because by assigning in the pause process the color $\Delta=8\omega(G')+1$ to $\delta_L$ and $\delta_R$ we have not ruled out the availability of color $\Pi$ for interval $\delta_{i+3}$. And since $\Delta$ is greater than both $L$ and $R$, $\chi_1=\chi'_1, \cdots, \chi_i=\chi'_i$ and the overlapping constraints for $\delta_{i+3}$ are the same in both processes, then $\Pi$ is the least color available for $\delta_{i+3}$. So we have again that $\chi_{i+3} \leq \chi'_{i+3}$.

\item If in the non-pause process it is the case that $R<\Pi$ and $L<\Pi$, then we know for sure that $\Pi$ is available because $\chi'_{i+3}=\Pi$. We also know that since  colors $L$ and $R$ are not assigned anymore to $\delta_L$ and $\delta_R$ they may be available for $\delta_{i+3}$ depending on its neighbors. In any case we will have that $\chi_{i+3}$ is going to be the least available color from the set $\{L,R,\Pi\}$. So again $\chi_{i+3} \leq \chi'_{i+3}$.

\item If $R<\Pi<L$, then as before $\Pi$ is available for $\delta_{i+3}$ and $L, R$ may or may not be available depending on the constraints. In any case $\chi_{i+3}$ is going to be the least available color from the set $\{R,\Pi\}$. So again $\chi_{i+3} \leq \chi'_{i+3}$.

\end{itemize} 

\end{itemize}

\underline{Induction Hypothesis}: $\chi_l \leq \chi'_l$, $\forall l, \, \, i+3 \leq l \leq j$

~\

\underline{Induction Step}: We want to prove that $\chi_{j+1} \leq \chi'_{j+1}$. Assume that in the non-pause process $\chi'_{j+1}=\Pi$. Suppose for the sake of contradiction that $\chi_{j+1}>\Pi$. This means that in the pause process there exists a neighbor $\delta_h$ of $\delta_{j+1}$ that appeared after $\delta_i$ with $h<j+1$ and was assigned the color $\chi_{\delta_{h}}=\Pi$. Now since $h \leq j$ and $\chi'_{\delta_h} \neq \Pi$, from induction hypothesis we have that $\chi_{\delta_{h}}$ is strictly less than $\chi'_{\delta_{h}}$ ($\chi_{\delta_{h}} < \chi'_{\delta_{h}}$). But this leads to a contradiction because of the following reason: Let $\chi'_{\delta_{h}}>\Pi$. This means that in the non-pause process $\Pi$ was not available for $\delta_h$, which means that $\delta_h$ had a neigbor $\delta_g$ with $g<h$ such that $\chi'_{\delta_g}=\Pi$. But from induction hypothesis this means that $\chi_{\delta_g}<\Pi$. So we discovered another interval $\delta_g$ for which $\chi_{\delta_g}$ is strictly less than $\chi'_{\delta_g}$ ($\chi_{\delta_g} < \chi'_{\delta_g}$). Someone can continue this process and either discover an infinite number of intervals appearing after $i$ that satisfy this strictness in the inequalities (which is a contradiction because the process is finite), or discover an interval $\delta_f$ for which $f < i$ and $\chi_{\delta_f} < \chi'_{\delta_f}$ which is also a contradiction since we know that $\chi_1=\chi'_1, \cdots, \chi_i=\chi'_i$. So the assumption that $\chi_{j+1}>\Pi$ is false and we have that $\chi_{j+1} \leq \chi'_{j+1}$.

\section{Circular arc graph online coloring using $FF$}

Let us now turn our attention to circular arc graph online coloring using \textit{FF}. We have the following

~\

\textbf{Proposition 1}: A Circular arc graph $G$ with minimum clique size 1 and maximum clique size $w(G)$ can be colored online by \textit{FF} with at most $8w(G)+1$ colors.
 
\begin{figure}[htb]
\centering 
\includegraphics[scale=0.4]{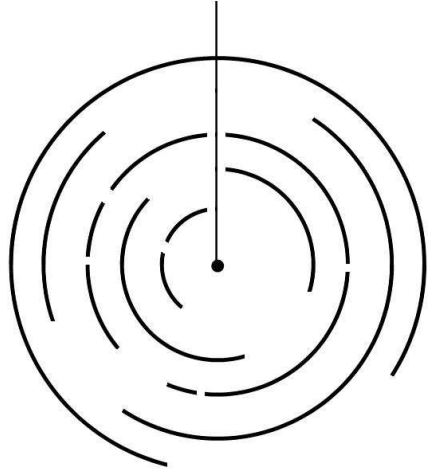}
\caption{A circular arc graph $G$ with minimum clique of size 1.}
\label{G}
\end{figure}

\textbf{Proof}: Suppose $G$ is a circular arc graph containing an arc $\delta$ such that if removed $G$ can be converted to an interval graph\footnote{A different definition of $G$ would be that $G$ has a minimum clique of size 1.}. Assume that $w(G)$ is the maximum clique size of $G$ and there exist $m$ arcs in $G$, none of which extends $360^{\circ}$ or more. We can see one such $G$ in Figure \ref{G}.

Suppose $G''$ is the interval graph that results if we "cut", $G$ in $\delta$ and loosely speaking "unfold" it in the cut. It is obvious that for $G''$ it holds $w(G'')=w(G)$. Also $G''$ has $m+1$ intervals because when arc $\delta$ was "cut" it split into two intervals, say, $\delta_L$ and $\delta_R$. It is obvious that $G''$ is identical with the interval graph $G'$ of the introduction, so in the rest of the note we will refer to $G'$ as the graph resulting  from cutting and unfolding the circular arc graph $G$. We also have to mention that the restriction we posed in the introduction that intervals $\delta_L$ and $\delta_R$ should not overlap is equivalent with the fact that in $G$ arc $\delta$ does not extend $360^{\circ}$ or more.

We want to prove that the maximum number of colors that $FF$ will use to color online $G$ is at most $8w(G)+1$ colors. For this and for our analysis to be easier we think of the process of online coloring the circular arc graph $G$ as the following equivalent\footnote{Two online colorings are equivalent when for the same sequences of intervals they produce the same chromatic sequences.} process of online coloring graph $G'$: While coloring $G$, when $FF$ is about to assign a color to arc $\delta$, it goes through the following discrete steps: 1st step is to locate the left neighbors of $\delta$ and find the color that can be assigned to $\delta$ if $\delta$ was overlapping only with these left neighbours - assume that this color is $L$. 2nd step is to do the same for the right neighbors and assume that it finds $R$. But since arc $\delta$ should get a unique color, finally, the 3rd step would be for $FF$ to assign the least available color $\chi_{\delta}$, greater or equal than both $L$ and $R$, i.e., $\chi_{\delta}\geq max\{L,R\}$.

Now while coloring $G'$, these steps are more clear. Since in $G'$ arc $\delta$ breaks into two intervals $\delta_L$ and $\delta_R$, we can say that there exists an online coloring of $G'$ in which $\delta_L$ and $\delta_R$ appear sequentially one after the other (say $\delta_L$ comes first and $\delta_R$ follows), $\delta_L$ has to get color $L$ (1st step) and $\delta_R$ has to get color $R$ (2nd step). Suppose now that someone pauses the process and finds the appropriate smaller available color $\chi_{\delta}$ that can be assigned both in $\delta_L$ and in $\delta_R$ such as $\chi_{\delta} \geq max\{L,R\}$, and then allows the process to continue. In this way he have mimicked the process of online coloring circular arc graph $G$ with the process of online coloring interval graph $G'$.

\begin{figure}[h]
\centering 
\includegraphics[scale=0.5]{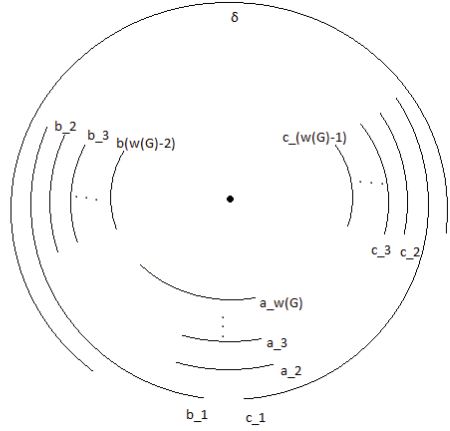}
\caption{A construction of a circular arc graph $G$ that forces $FF$ to use $8w(G)+1$ colors in an online coloring.}
\label{construction}
\end{figure}

So using the above analysis every online coloring by $FF$ of an arbitrary sequence of the arcs of $G$, $\Sigma=<\delta_1,\delta_2,...,\delta_{j-1},\delta,\delta_{j+1},...,\delta_m>$, can be achieved as follows: While $FF$ online colors sequence $\Sigma$, when arc $\delta$ appears, $FF$, before assigning a color, breaks\footnote{It is as if in the background $FF$ works with the sequence $\Sigma'=<\delta_1,\delta_2,...,\delta_{j-1},\delta_L,\delta_R,\delta_{j+1},...,\delta_m>$ in order to decide the unique color arc $\delta$ should receive.} $\delta$ into two intervals $\delta_L$ and $\delta_R$. It assigns to $\delta_L$ a color (say $L$) and to $\delta_R$ a color (say $R$). But since arc $\delta$ must receive a uniqe color, finally, $FF$ must assign at most the color $8w(G)+1$ to interval $\delta$. This is because there exists a case that the least available color for $\delta$ can't be any from the set $\{1,\cdots,8w(G)\}$. Consider for example the case where $\chi_{\delta_L}=8w(G)-1$, $\chi_{\delta_R}=8w(G)$ and $\delta$ overlaps with an arc that has assigned previously the color $8w(G)$ (see the example below and Figure \ref{construction} for such a construction). Clearly in this case the minimum color that satisfies both constraints of $\chi_{\delta_L}$ and $\chi_{\delta_R}$ is $8w(G)+1$. Using now Lemma 1, $FF$ is sure that no other arriving arc after $\delta$ will get a color bigger than $8w(G)$, and thus bigger than $8w(G)+1$, which completes the proof of Proposition 1.

~\

Before moving to the general case let us give the above-mentioned construction that forces $FF$ to assign the extra color $8w(G)+1$. In Figure \ref{construction} we see a circular arc graph $G$ with minimum clique size 1 and maximum clique size $w(G)$ (for example, the arcs $\{c_1, a_2,\cdots,a_{w(G)}\}$ is such a maximum clique). Consider now the following order of appearance of the arcs in an online coloring of $G$: 

$\Sigma=<b_1, c_1, a_2, a_3,\cdots,a_{w(G)},c_2,c_3,\cdots,c_{w(G)-1},b_2,b_3,\cdots,b_{w(G)-2},\delta>$. The chromatic sequence produced is $X=<1,1,2,3,\cdots,w(G),2,3,\cdots,w(G)-1,2,3,\cdots,w(G)-2,w(G)+1>$ and this is because when $FF$ is ready to assign a color to arc $\delta$ it first brakes $\delta$ into $\delta_L$, $\delta_R$ and assignes $\chi_{\delta_L}=w(G)-1$ and $\chi_{\delta_R}=w(G)$, but when it trys to merge $\delta_L$ and $\delta_R$ to form $\delta$ no color from the set $\{1,\cdots,w(G)\}$ is available due to overlaping constraints. So the only solution is for color $8w(G)+1$ to be introduced and assigned to $\delta$.

\section{The general case}

In the general case graph $G$ would have a maximum clique of size $w(G)$ and a minimum clique of size $K \leq w(G)$, which if we cut we can convert $G$ into an equivalent interval graph $G'$. We will prove that in this case $FF$ needs at most $8w(G)+K$ colors to color online $G$. The proof would be by induction on $j \leq K$.

~\

\textbf{Theorem}: Circular arc graph $G$ can be colored online using $FF$ with at most $8w(G)+K$ colors.

\textbf{Proof}: Induction on $j \leq K$.

\underline{Base}: $j=1$,  then $G$ has a minimum clique of size 1 and it holds from Proposition 1.\\

\underline{Induction Hypothesis}: Suppose that it holds for $j$.

~\

\underline{Induction Step}: Assume $G$ is a circular arc graph with minimum clique of size $j+1$. Assume that in an online coloring of $G$ the first $j$ arcs of this maximal clique have appeared, and $\delta$ is the last arc that needs to appear. Until now we have from the induction hypothesis that no more that $8w(G)+j$ colors have been used. We have to show that from now on at most $8w(G)+j+1$ colors will be used.

As before we can assume that arc $\delta$ can break into two intervals $\delta_L$ and $\delta_R$. So the resulting graph is a circular arc graph with minimal clique of size $j$ for which graph the induction hypothesis holds. The process of coloring arc $\delta$ can be replaced by the following 3 steps: 1) $FF$ finds the available color from the left constraints, say it is $L$, 2) $FF$ finds the available color from the right constraints, say it is $R$ and 3) $FF$ finds the minimum available color greater than or equal to $L$ or $R$ and colors $\delta$.

For this case it is also possible to prove a lemma similar to lemma 1 and a proposition similar to proposition 1 and finally applying these prove that the hypothesis also holds for $j+1$.

~\

\textbf{Corollary}: If $G$ is a circular arc graph with maximal clique $w(G)$, then it can be online colored by $FF$ with at most $9w(G)$ colors.

\textbf{Proof}: We apply the theorem for $K=w(G)$.




\begin{thebibliography}{999}

\bibitem{Raman} Raman, R. Chromatic Scheduling. In {\em Ph.D. thesis}; University of Iowa, July 2007

\bibitem{Golumbic} Golumbic, M.C.
In {\em Algorithmic Graph Theory and Perfect Graphs},
Academic Press, New York, 1980

\bibitem{Tucker} Tucker, A. An Efficient Test for Circular Arc Graphs.
{\em SIAM Journal on Computing} {\bf 1980}, {\em 9, 1}, 1-24.

\bibitem{Damaschke} Damaschke, P. Paths in Interval Graphs and Circular Arc Graphs. {\em Discrete Mathematics} {\bf 2008}, {\em 112}, 49-64.

\bibitem{ISGCI} ISGCI, Information System on Graph Classes and their Inclusions, 
http://wwwteo.informatik.uni-rostock.de/isgci/index.html 

\bibitem{Gyarfas} Gyarfas, A; Lehel, J. On-Line and First Fit Colorings of Graphs. {\em Journal of Graph Theory} {\bf 1988}, {\em 12, 2}, 217-227.

\bibitem{Kierstead} Kierstead, H. A.; Qin J. Coloring Interval Graphs with First-Fit. {\em Discrete Mathematics} {\bf 1995}, {\em 14, 1-3}, 47 - 57
 
\bibitem{Marathe} Marathe. M. V.; Hunt, H. B.; Ravi, S. S. Efficient Approximation Algorithms for Domatic Partition and On-line Coloring of Circular Arc Graphs. {\em Discrete Applied Mathematics} {\em 1996}, {\em 64, 2}, 135 - 149. 

\bibitem{Slusarek} Slusarek, M. Optimal On-Line Coloring of Circular Arc Graphs. {\em Informatique Theorique et Applications} {\bf 1995}, {\em 29, 5}, 423 - 429.


\end{thebibliography}
\end{document}